\documentclass[5p]{elsarticle}
\usepackage{amsmath,amssymb}
\usepackage[amssymb,thinqspace,thinspace]{SIunits}
\usepackage{color}
\usepackage{graphicx}
\usepackage{tabularx}
\usepackage{lineno}

\journal{Met Trans A}

\begin{document}

\begin{frontmatter}

\title{The Kinetics of Primary Alpha Plate Growth in Titanium Alloys}

\author[IC]{Abigail K. Ackerman}
\author[IC]{Alexander J. Knowles}
\author[OX]{Hazel M. Gardner}
\author[OX]{Andr\'{e} A. N. N\'{e}meth}
\author[IC]{Ioannis Bantounas}
\author[RR]{Anna Radecka}
\author[OX]{Michael P. Moody}
\author[OX]{Paul A.J. Bagot}
\author[OX]{Roger C. Reed}
\author[RR]{David Rugg}
\author[IC]{David Dye}

\address[IC]{Department of Materials, Royal School of Mines, Imperial College London, Prince Consort Road, London, SW7 2BP, UK}
\address[OX]{Department of Materials,University of Oxford, Parks Road, Oxford, OX1 3PH, UK}
\address[RR]{Rolls-Royce plc., Elton Road, Derby, DE24 8BJ, UK}

\begin{abstract}
The kinetics of primary $\alpha$-Ti colony/Widmanst\"{a}tten plate growth from the $\beta$ are examined \textcolor{black}{in Ti-6246}, comparing a simple quasi-analytical model to experiment. The plate growth velocity depends sensitively both on the diffusivity $D(T)$ of the rate-limiting species and on the supersaturation around the growing plate. These result in a maxima in growth velocity around $40\usk\kelvin$ below the transus, once sufficient supersaturation is available to drive plate growth.  In Ti-6246, the plate growth velocity was found to be around $0.32\usk\micro\meter\usk\minute^{-1}$ at $850\usk\celsius$, which was in good agreement with the model prediction of $0.36\usk\micro\meter\usk\minute^{-1}$. The solute field around the growing plates, and the plate thickness, was found to be quite variable, due to the intergrowth of plates and soft impingement. This solute field was found to extend to up to $30\usk\nano\meter$, and the interface concentration in the $\beta$ was found to be around $6.4\usk$ at.\% Mo. It was found that increasing O content \textcolor{black}{from 500 to 1500wppm} will have minimal effect on the plate lengths expected during continuous cooling; in contrast, Mo approximately doubles the plate lengths obtained for every 2 wt.\% Mo reduction. Alloys using V as the $\beta$ stabiliser instead of Mo are expected to have much faster plate growth kinetics at nominally equivalent V contents.  These findings will provide a useful tool for the integrated design of alloys and process routes to achieve tailored microstructures.
\end{abstract}


\end{frontmatter}

\section{Introduction}
\label{Intro}

Titanium alloys are widely used in aerospace gas turbines owing to their unrivalled specific fatigue-allowable strengths \cite{Titanium,Boyer1995}. In the cold sections of the engine, e.g. the fan, Ti-6Al-4V is used in large forgings (discs) and plate and bar (blades), primarily owing to its relatively low density and ease of forging and machinability \cite{Boyer2007}. In the highest temperature sections of the compressor, near-$\alpha$ alloys are used due to their favourable creep performance, as the solute diffusivity is lowest in the hexagonal close packed alpha phase ~\cite{Murray1990,Weiss1999a}. Often, strongly $\beta$ stabilised \textcolor{black}{$\alpha+\beta$} alloys such as Ti-17 and Ti-6Al-2Sn-4Zr-6Mo are also used at intermediate temperatures in the compressor, since they retain their strengths to higher temperatures than Ti-6Al-4V \cite{Boyer1995}. However, their elevated Mo content reduces forgeability and machinability and increases the density. Nevertheless Mo provides the opportunity to slow down the kinetics of formation of the $\alpha$ phase \cite{Semiatin1996}, enabling basketweave multi-variant primary $\alpha$ microstructures to be attained with a high fraction of secondary $\alpha_s$, which result in a very strong, fatigue resistant alloy \cite{Qiu2014a}.

In the scheme of Ti-6246 processing, the cooling rate from the $\beta$ can be controlled in order to ensure that the primary $\alpha$, which nucleates at the prior beta grain boundaries, fills the prior $\beta$ grains. This is then followed by an ageing heat treatment to precipitate out the secondary $\alpha$ from solution the finest possible plate thickness. These three parameters - the cooling rate from the $\beta$, the secondary ageing temperature and ageing time, enable the desired strength-toughness balance to be chosen according to the application.  However, a process model of the primary $\alpha$ growth has not been presented, and therefore such optimisation has historically been performed empirically \cite{Titanium}. 

Classical quasi-analytical models for $\alpha$ precipitation have been developed by Semiatin and co-workers \cite{Semiatin2003}.  They used the Ivantsov solution \cite{Ivantsov1947} for parabolic growth in an infinite medium, which was developed for the cases of 2D paraboloids by Horvay and Cahn\cite{Horvay1961}, taking the spherical solution to examine globular precipitate growth during the globularisation heat treatment step in Ti-6Al-4V \cite{Warwick2011,Kherrouba2016}. They found that the growth of globular primary $\alpha$ is well defined by a diffusion controlled process, based upon diffusion of aluminium and vanadium. However the constant radius solution as outlined in their work underestimates the growth rate of primary $\alpha$. The model can also be applied to predict the temperature at which primary $\alpha$ ceases forming and secondary $\alpha$ growth begins. Therefore, a successful scheme exists for the modelling of $\alpha$ precipitation and growth, but has not been applied to the growth of lath-shaped primary $\alpha$ from the grain boundaries. \textcolor{black}{Additional investigations have been completed by Gao et al.~\cite{Gao2017}, where the size of primary globular $\alpha$ was successfully predicted and compared to experimental data, based upon the work of Semiatin et al.\textcolor{black}{~\cite{Semiatin2003}}. Appolaire et al.~\cite{Appolaire2005} developed a kinetic model based upon nucleation and growth laws, comparing it to a metastable $\beta$ titanium alloy, with good agreement.}

Colony and Widmanst\"atten microstructures are produced by varying the cooling rate from the $\beta$ phase field; a slow cooling rate will result in a colony $\alpha$ microstructure, and a rapid cooling rate will result in a basketweave microstructure~\cite{Ahmed1998,Lutjering1998, Gao2017}.
The $\alpha$ phase grows in accordance with the Burgers orientation relationship $\{0001\}\parallel\{110\}$, \textcolor{black}{($[11\Bar{2}0]\parallel[1\Bar{1}1]$),} which provides an interface with the lowest dislocation energy \cite{Plichtai1980}. Nucleation commonly occurs from the prior $\beta$ grain boundary, as titanium lacks ceramic inclusions or other heterogeneous nucleation sites. At slow cooling rates, a grain boundary film of $\alpha$ often forms, particularly where the recrystallisation texture gives rise to $\beta$ grain boundaries where an $\alpha$ crystallographic variant can be found that satisfies the Burgers orientation relationship on both sides of the grain boundary. This has been suggested to result in the survival of the $\alpha$ texture on cycling above the $\beta$ transus~\cite{Stanford2004}. Grain boundary $\alpha$ is commonly held to result in poor fatigue performance. Where such a grain boundary film forms, growth of the Widmanst\"{a}tten $\alpha$ colony then occurs from the grain boundary $\alpha$.  At higher cooling rates, a basketweave microstructure can be formed~\cite{Lee2007}, where the growing $\alpha_p$ plates branch, forming different crystallographic variants~\cite{Tong2017,Bhattacharyya2003a,Furuhara2001}.  The formation of a variety of $\alpha_p$ and $\alpha_s$ variants is likely to be important, because slip bands can be easily transmitted between similarly oriented grains~\cite{Joseph2018}, particularly if the $\alpha$ phase is rich enough in Al that low temperature ageing precipitates the $\alpha_2$ Ti$_3$Al phase~\cite{Radecka2016,Lim1976,Neeraj2001,Williams1972}. Slip localisation can also be associated with other effects, such as hydrogen and stress corrosion embrittlement~\cite{Chapman2016}. 

From an industrial perspective and in the framework of Integrated Computational Materials Engineering (ICME), it would be useful for the industrial process engineer to possess a simple model for primary $\alpha$ precipitation that could act as a guide to optimise the cooling rate from the $\beta$ field.   Here we develop and test such a model.  The diffusion field around the primary $\alpha$ plates is examined by atom probe tomography (APT) and scanning transmission electron microscopy-energy dispersive x-ray (STEM-EDX) for the purposes of comparison, along with measurement of the growth rate from isothermal holding experiments.  The model is then used to develop processing maps to characterise the effect of cooling rate on the ability to fill the prior $\beta$ grain.

\section{Modelling}
\label{Modeling}

\subsection{Growth Velocity Model}
\begin{figure*}[t!]
\centering\includegraphics[width=1\linewidth]{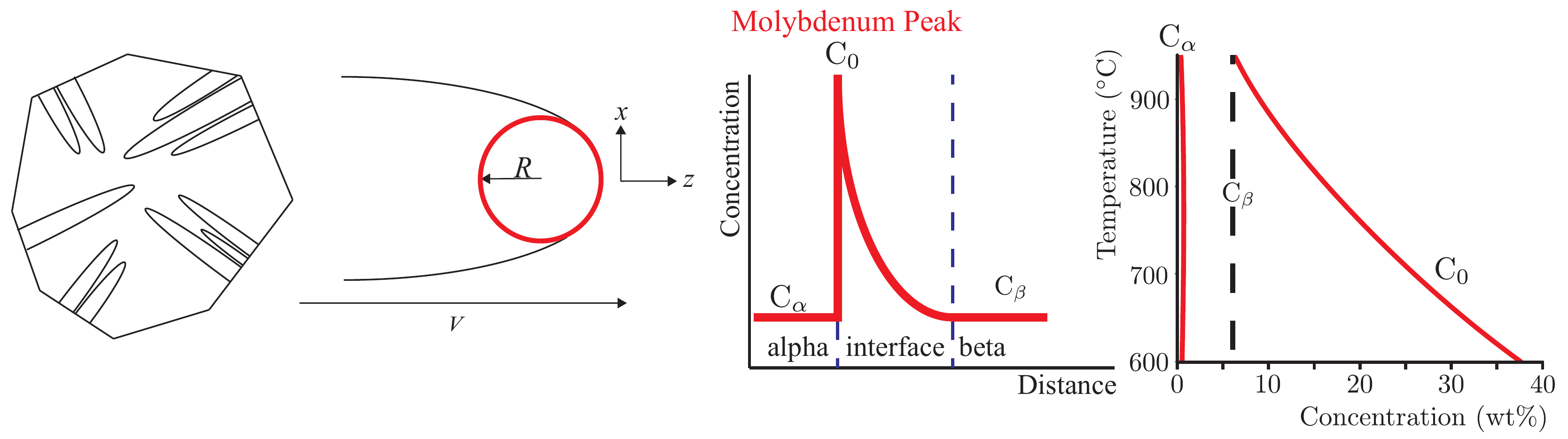}
\caption{(a) The model describes a growing $\alpha$ plate within a $\beta$ grain. The $\alpha$ plate is treated as a moving parabola, (b), with an x and z component, forming the basis of the 2-dimensional model. The parabola has a radius of curvature $\textit{R}$, growing with velocity $\textit{V}$. (c) $C_{\alpha}$ is the composition of the $\alpha$ phase, $C_\beta$ is the composition of the $\beta$ phase above the transus and $C_{0}$ is the composition at the interface. These values can be taken from the Ti-Mo phase diagram, (d), or by using a thermodynamic database software such as Thermocalc or Pandat. The resulting compositions can be used for modeling both the velocity of the growing $\alpha_p$ plate and the diffusion field of Mo away from the interface.}\label{fig:growthschem}
\end{figure*}

Ivantsov \cite{Ivantsov1947}, Zener~\cite{Zener1949} and Horvay and Cahn \cite{Horvay1961} first developed a solution for the thermal diffusion problem around a growing paraboloid, making the assumption that the compositional matrix around the growing precipitate is constant\footnote{The presentation here is largely unchanged from that obtainable by following through the literature~\cite{Ivantsov1947,Horvay1961,Christian}, but is presented in a single treatment here for clarity; the interested reader should be aware that there is a typographical error in Equation 1 of ~\cite{Trivedi1970}}. However, for isothermal transformation, the interfacial curvature and kinetics must be taken into account. Bosze and Trivedi,~\cite{Bosze1975,Trivedi1970} and Liu and Chang \cite{Liu1997} developed the approach, using the Ivantsov equation with the Trivedi modification and applying this to carbon composition close to an interface with curvature. In the present work, the growing alpha plate is treated as a paraboloid with tip growth velocity \textit{V}, Figure~\ref{fig:growthschem}.

The radius of curvature of the tip $R$, diffusivity $D$ and velocity are inter-related through the Peclet number, $\textit{P}$, Equation~\ref{eq:vel}. The Peclet number is a dimensionless speed of advancement, obtained from Ivantsov's solution to the diffusion equation,~\ref{eq:peclet}: 

\begin{equation}
\label{eq:vel}
P = VR/2D(T)
\end{equation}

\begin{equation}
\label{eq:peclet}
\Omega = \sqrt{\pi P} e^P (1-\mathrm{erfc}\sqrt{P})
\end{equation}

where $\Omega$ is the supersaturation, Equation~\ref{eq:Omega}, which is a thermodynamic parameter of the system; $0\leq\Omega\leq1$.

\begin{equation} 
\label{eq:Omega}
\Omega(T) = \frac{C_0 - C_\beta}{C_0 - C_\alpha}
\end{equation}

Here, $C_\beta$ is the far-field $\beta$ composition (alloy composition), $C_\alpha$ is the interface concentration in the growing $\alpha$ lath, and $C_0$ is the solubility in the $\beta$ at the interface.  Solving Equation~\ref{eq:peclet} (numerically) for $\textit{P}$ gives a dependence of velocity on temperature $T$, through its effect on $\Omega$ and $D$. 

\subsection{Diffusion Model}
For validation of the growth velocity model, we compare the diffusion fields around the growing plates.  In the mathematical procedure first used by Aaron \emph{et al}. \cite{Aaron1970}, the growing primary $\alpha$ plate is treated as an isolated \textcolor{black}{lamella of tip} radius $R$ within an infinite $\beta$ matrix, in a diffusion-limited transformation governed by Fick's second law 

\begin{equation}
D\nabla ^2C = \partial C / \partial t \label{eq:diffusion}
\end{equation}

where \textit{D} is the diffusion coefficient in the matrix and \textit{C = C(r, t)} is the concentration field around the $\alpha$ precipitate in the $\beta$ matrix, with the boundary conditions:

\begin{equation}
\label{eq:11}
\begin{split}
C(r=R, t)&=C_0 \qquad  0<t\leq\infty\\
C(r, t=0)&=C_\beta \qquad  r\geq R\\
C(r=\infty,t)&=C_\beta \qquad  0\leq t\leq\infty 
\end{split}
\end{equation}
 
where r=R at the $\alpha/\beta$ interface.  The flux balance must be satisfied:

\begin{equation}
 \label{eq:12}
(C_\alpha - C_0)\frac{dR}{dt} = D\frac{\partial C}{\partial R}\Big|_{r=R}
\end{equation}

where $C_\alpha$ is the composition of the growing $\alpha$ phase.  For planar growth, Equation~\ref{eq:diffusion} becomes:

\begin{equation}
\label{eq:13}
C(x, t) - C_\beta = (C_0 - C_\beta)\frac{\mathrm{erfc}[(x/2)\sqrt{Dt}]}{\mathrm{erfc}(\lambda)}
\end{equation}

where $x=S$ at the $\alpha/\beta$ interface; $x$ is the Cartesian distance from the flat interface, Figure~\ref{fig:growthschem}. The exact solution for this boundary condition is:

\begin{equation}
S=2 \lambda \sqrt{Dt}
\end{equation}

where $\lambda$ is the solution to:

\begin{equation}
\sqrt{\pi}\lambda \exp\{\lambda^2\}\,  \mathrm{erfc} \lambda=-\Omega/4
\end{equation}
where the supersaturation is $\Omega$. Alternately, if the interface is assumed to be in a stationary steady-state, the following approximation can be used:

\begin{equation}
\label{eq17}
C(r,t)- C_\beta = (C_0 - C_\beta) \mathrm{erfc}\Big\{\frac{x-S}{2\sqrt{Dt}}\Big\}
\end{equation}

As $S=0$ when $t=0$, it can be said that

\begin{equation}
S = -\frac{2\Omega}{\sqrt{\pi}}\sqrt{Dt}
\end{equation}
Equations (4) to (11) can be arranged to obtain the concentration gradient away from the interface, dependent on the concentration in the $\alpha$ and $\beta$ phases and the temperature. 

\section{Experimental Procedure}
\label{Experimental}
Samples were prepared from a high pressure compressor disc of measured (by scanning electron microscope-energy dispersive x-ray SEM-EDX) composition Ti-6Al-2Sn-4Zr-6Mo (wt\%), corresponding to Ti-10.8Al-0.82Sn-2.13Zr-3.04Mo (at.\%), supplied by Rolls-Royce plc. Derby. A typical heat chemistry provided by Timet for this product is around 1050 wppm O.

\begin{figure}[t!]
\centering\includegraphics[width=1.0\linewidth]{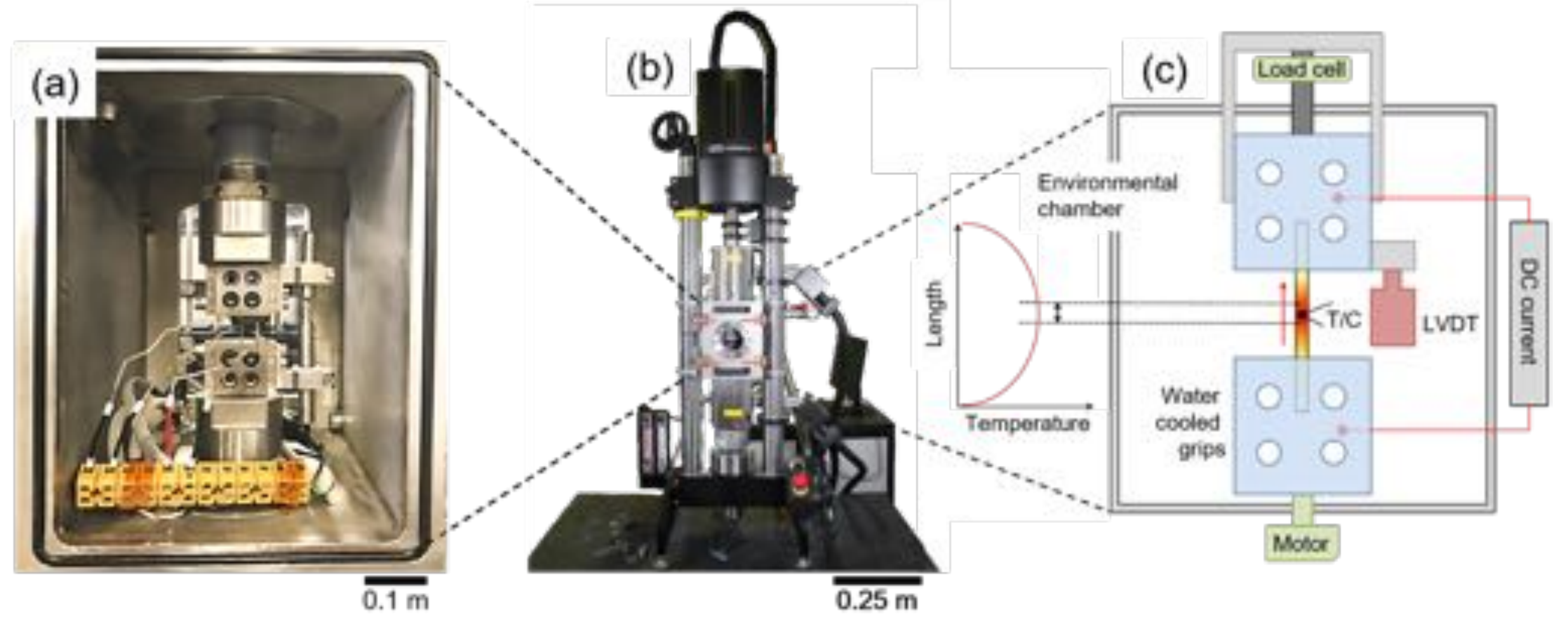}
\caption{Electro-Thermal Mechanical Testing (ETMT) was used to achieve rapid heating and cooling rates. A `matchstick' sample of Ti-6246, \textcolor{black}{$50\times2\times2\usk\milli\meter$}, was held between water cooled steel grips. A Pt-Rh control thermocouple was attached at the central point of the sample and the sample heated by resistive heating.}\label{fig:etmt}
\end{figure}

\begin{figure}[t!]
\centering\includegraphics[width=1.0\linewidth]{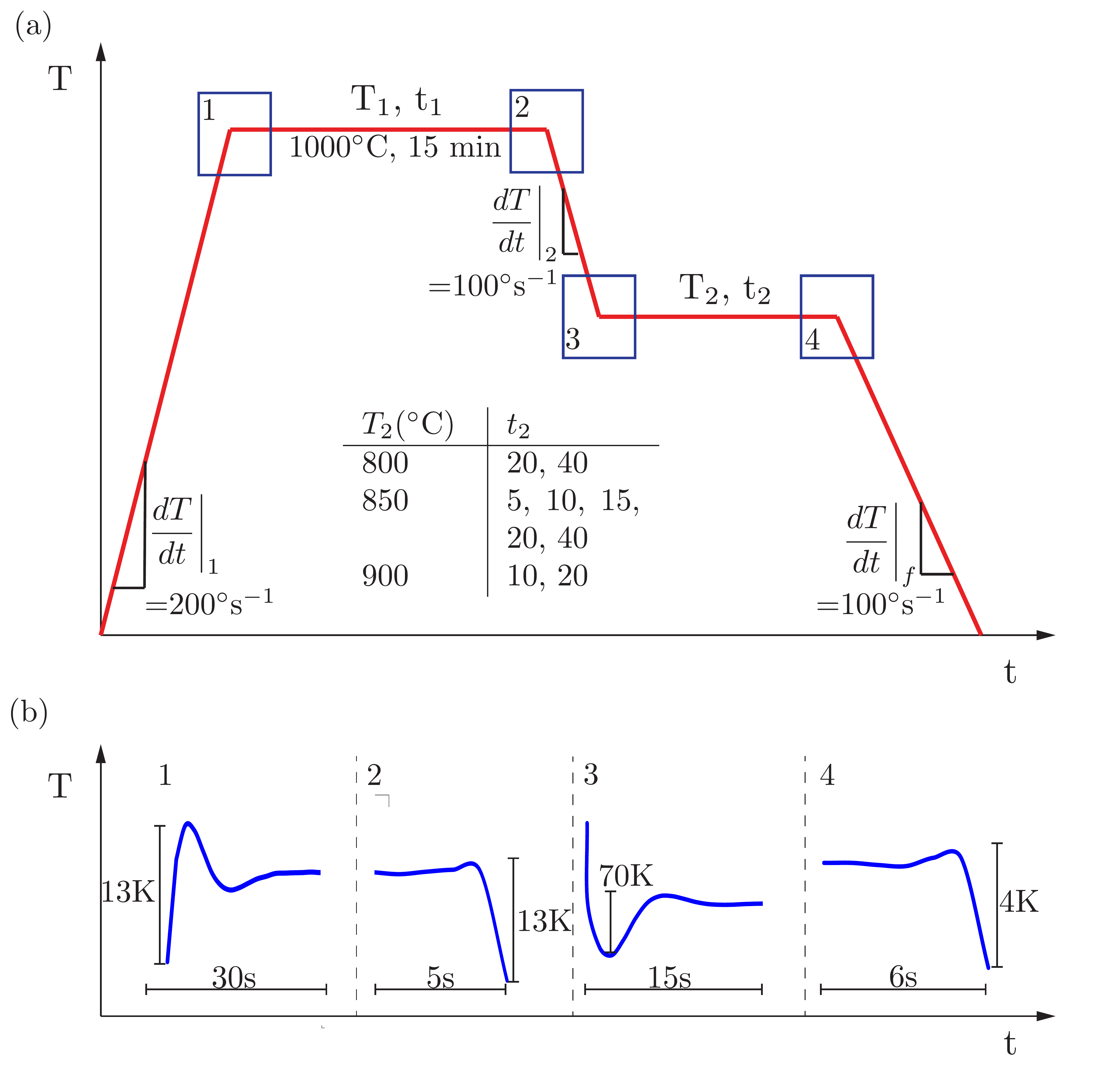}
\caption{Schematic of the applied heat treatments during ETMT testing to measure the kinetics of Widmanst\"{a}tten plate growth. Inset table shows final temperatures and hold times for the samples.~(b) shows the effectiveness of the temperature control at transition points. }\label{fig:thermal}
\end{figure}

\begin{figure}[h]
\centering\includegraphics[width=1.0\linewidth]{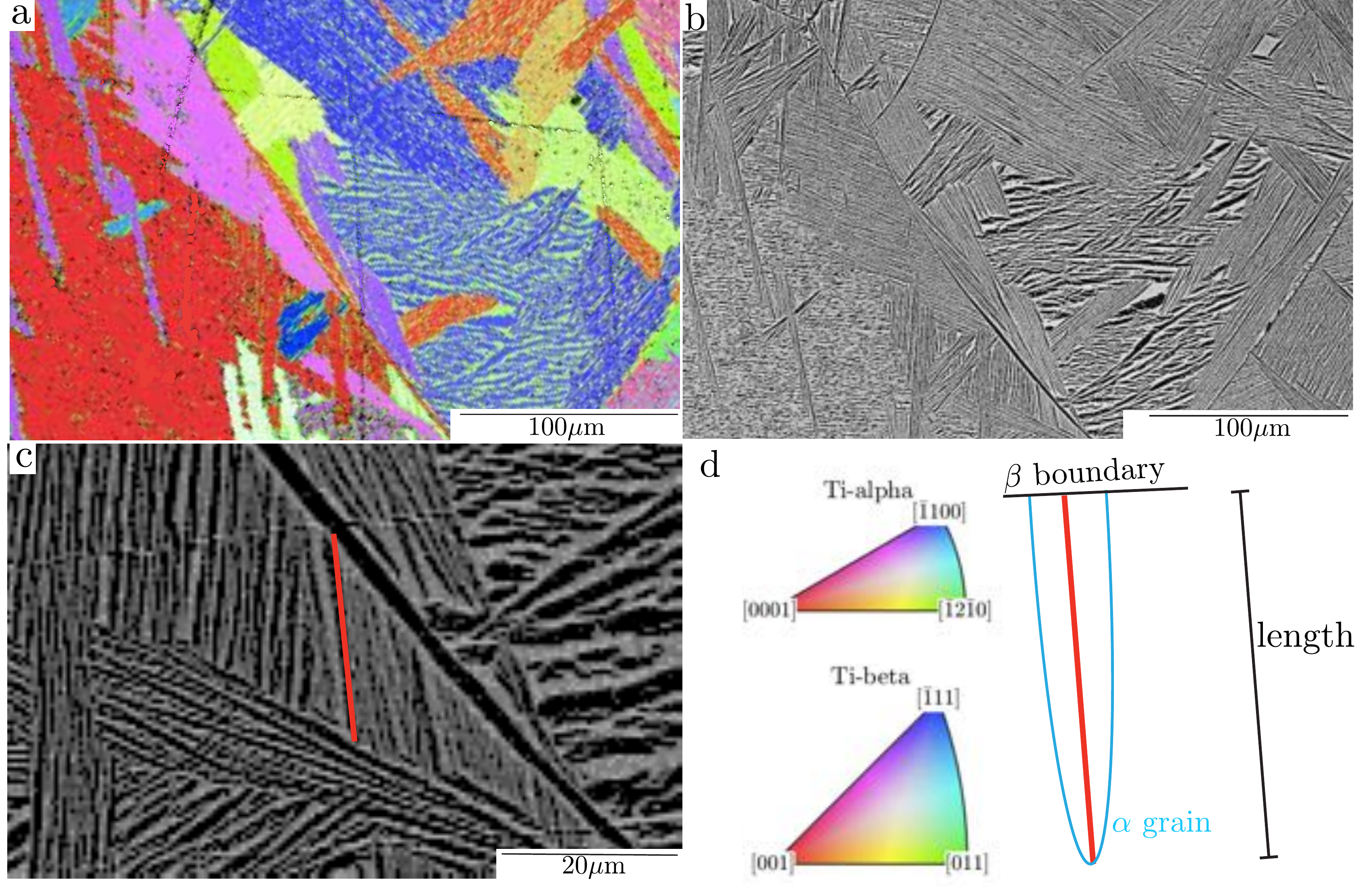}
\caption{To obtain lengths for velocity calculations, the following procedure was completed: (a) EBSD was collected to find an area where the $\alpha$ laths were normal to the viewing plane, in the $(0002)_\alpha || \{110\}_\beta$ direction, (b) the area was then viewed using backscattered diffraction and matched with the EBSD image to ensure the same grain was measured, (c) higher resolution backscatter imaging was done to view the grains, (d) the length was measured along a minimum of 20 $\alpha$ grains, from the grain boundary to the tip. The $\alpha$ width of these grains was also measured using ImageJ.}\label{fig:measure}
\end{figure}

A 5 kN Instron electro-thermal mechanical testing (ETMT) machine was used to rapidly change the temperature of the material in order to determine the isothermal growth rate of Widmanst\"{a}tten $\alpha$ plates, Figures~\ref{fig:etmt}--\ref{fig:thermal}. $2\times 2\times 50\usk\milli\meter$ samples of as-received Ti-6246 were tested in backfilled argon under load control at 900, 850 and $800\usk\celsius$, at hold times of 5--$50\usk\minute$. Temperature measurement during testing was achieved by spot-welding a K-type thermocouple to the central portion of the specimen. Consistent with the design of the ETMT machine,
specimens were heated using electrical resistance by passing a direct current through the specimen. \textcolor{black}{This creates a parabolic temperature profile controlled by a thermocouple placed at the (hottest) mid-point, Figure 2; care must therefore be taken in sectioning}. After testing, the samples were sectioned at the spot weld and prepared using standard metallographic procedures. Electron backscatter diffraction (EBSD) was performed using either a Zeiss Sigma 300 SEM or Zeiss Auriga FEG-SEM. An Oxford Instruments HKL eFlash EBSD detector was used to identify the orientation of the grains to ensure that the primary $\alpha$ laths that were measured had an orientation where $(0002)_\alpha || \{110\}_\beta$ was normal to the sample plane, to avoid stereological effects in the measurement of plate length. \textcolor{black}{A similar morphology of primary $\alpha$ was measured each time to ensure  consistancy between measurements.} Images were then collected in backscatter mode at $8\usk\kilo\volt$ and at $5\usk\milli\meter$ working distance in order to measure the plate growth lengths, Figure~\ref{fig:measure}, and measurements taken using the ImageJ open source software package. 

\begin{figure}[t!]
\centering\includegraphics[width=0.8\linewidth]{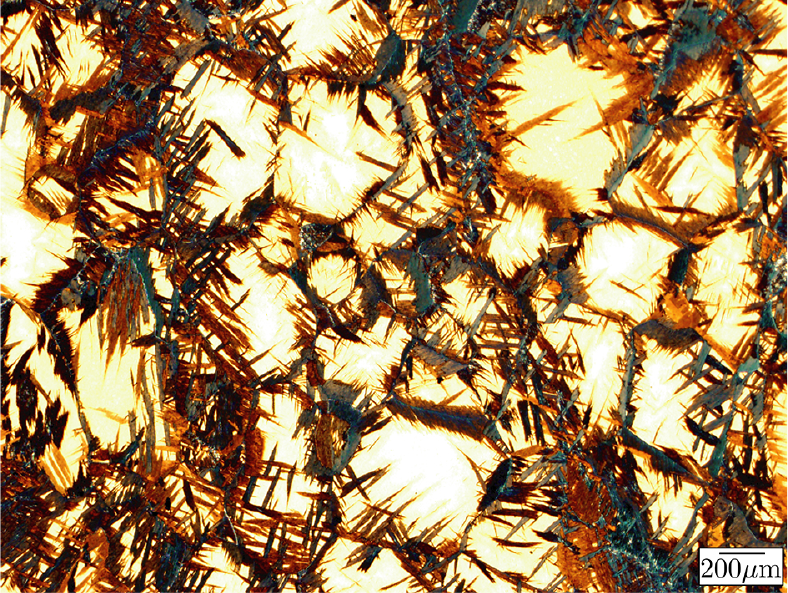}
\caption{A light micrograph taken using cross-polarising filters showing the microstructure of the alloy after being treated at $960\celsius$, cooled at  $7\celsius\usk\minute$$^{-1}$ to $800\celsius$ and water quenched manually. \textcolor{black}{The sample was etched for 5$\usk\second$ using Kroll's reagent (100$\usk\milli\litre$ H$_{2}$O, 6$\usk\milli\litre$ HNO$_{3}$, 3$\usk\milli\litre$ HF). $\alpha$ laths appear dark whilst the $\beta$ phase appears light. This provided primary $\alpha$ laths devoid of secondary $\alpha$ for the FIB lift out of TEM and APT samples.}}\label{optical}
\end{figure}

Transmission electron microscopy and atom probe tomography specimens were prepared from the same sample, in order to measure the diffusion profiles around the $\alpha_p$ plates. A $10\times 10 \times 10\usk\milli\meter$ sample was heat treated for $30\usk\minute$ at $960\celsius$ then cooled at $7\celsius\usk\minute$$^{-1}$ to $800\celsius$, followed by manual water quenching, Figure~\ref{optical}. Samples for transmission electron microscopy (TEM) were first investigated using a Zeiss Auriga FEG-SEM fitted with an Oxford Instruments HKL eFlash EBSD detector to ensure that the \textcolor{black}{$\{0001\}_\alpha|| \{110\}_\beta$} directions were contained with the foil plane.  Foils and atom probe needles were then prepared using an FEI Helios NanoLab 600 DualBeam system equipped with an Omniprobe$^\text{TM}$ using standard procedures~\cite{Thompson2007, Mayer2007}. Transmission electron microscopy was then performed using a JEOL 2100F 200kV, with use of an Oxford Instruments EDS detector and Aztec software. Atom probe tomography was performed using a LEAP 5000XR in laser mode with a laser energy of 40$\usk\pico\joule$, at a temperature of 50$\usk\kelvin$ and a pulse rate of 200$\usk\kilo\hertz$. APT data was reconstructed using the IVAS software which is based on the conventional reverse projection algorithm \cite{Gault2011}.

\section{Results}
\label{results}
\subsection{Velocity of Primary Alpha Growth}

\begin{figure}[t!]
\centering\includegraphics[width=1.0\linewidth]{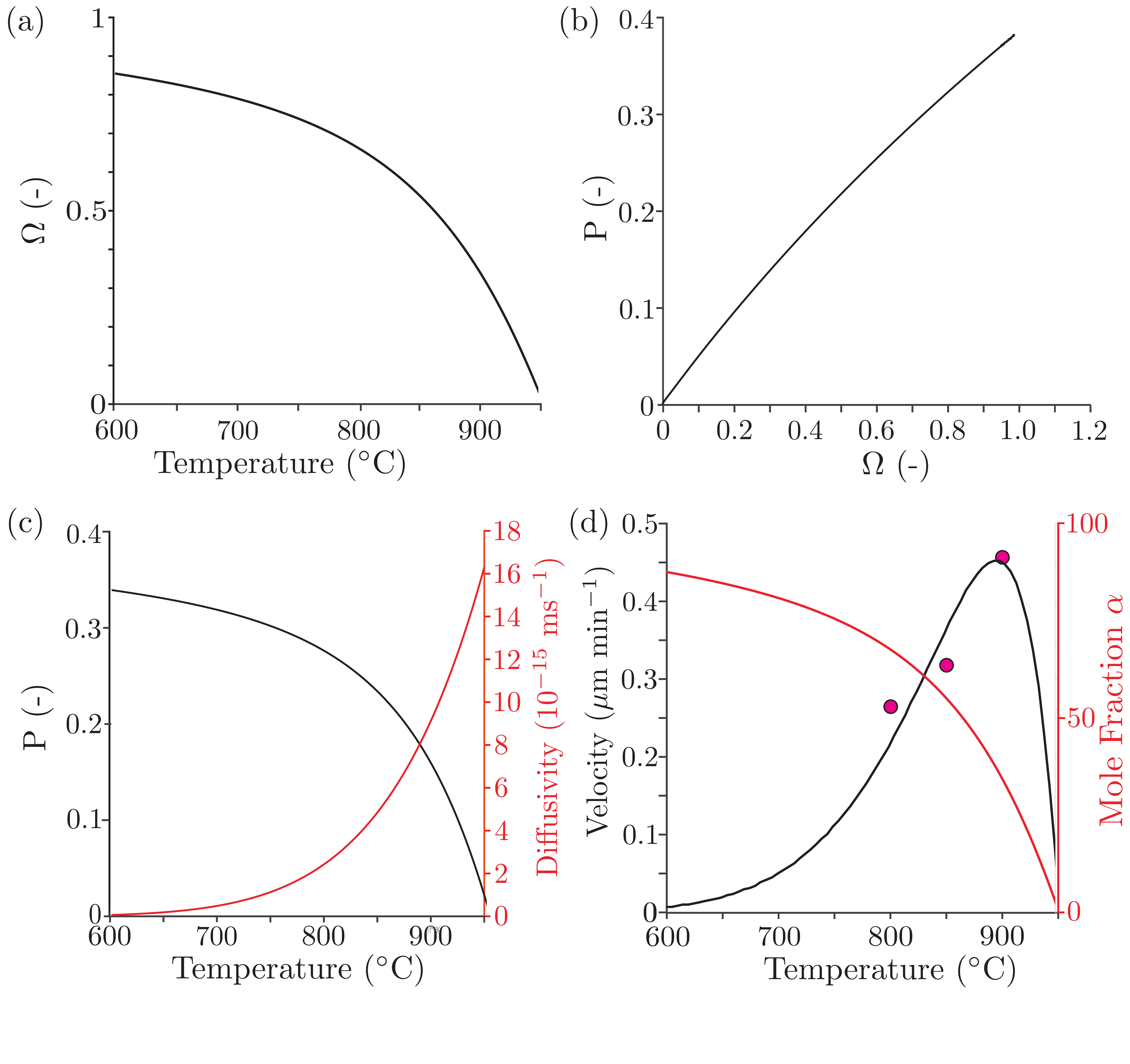}
\caption{Modelling results (a) Effect of temperature on supersaturations $\Omega$ and $k$, (b) effect of supersaturation $\Omega$ on dimensionless velocity $P$, (c) relationship of temperature to diffusivity of Mo in $\beta$-Ti and $P$. Finally (d) relationship of calculated plate tip velocity $V$ to temperature, and the prediction equilibrium $\alpha$ mole fraction. The points are the measured growth velocities, Figure~\ref{fig:vels-meas}.}\label{fig:peclet}
\end{figure}

The Pandat-predicted (Computherm, LLC) equilibrium phase compositions and corresponding thermodynamic parameters were obtained, for Ti-6246 containing 1050 ppmw O. From these, the supersaturation, $\Omega$, and partition coefficient, $k$, were derived. It can be observed that close to the transus, at high temperatures, $\Omega$ is small, and increases as the temperature drops, Figure~\ref{fig:peclet}. This is a simple consequence of the phase diagram \textcolor{black}{~\cite{Murray1990}}; at the transus, $C_0=C_\beta$, and so there is no chemical activity `push' from supersaturation to diffuse away rejected solute. In turn, when $\Omega$ is large, the Peclet number is large, and so, since the growth velocity is proportional to $P$, the supersaturation would result in higher growth velocities at lower temperatures.  The plate tip radius $R$ used in the model has been inferred experimentally from the observed plate thickness ($2R$) of $\usk0.75\usk\micro\meter$; it may be that this is temperature-dependent, but at this stage in the development of the theory, this aspect of plate growth is not captured. The growth velocity $V$ is also proportional to the diffusivity $D$, which has a slightly stronger temperature dependence than that of $P$. Since $V=2DP/R$, the velocity term is a consequence of both the ability of the matrix to diffuse the solute hill away from the interface and the supersaturation; at low temperatures growth is inhibited due to the diffusivity, whereas at high temperatures $P(\Omega)$ is small. Hence there is a maximum in the growth velocity at around $870\celsius$.

\begin{figure}[t!]
\centering\includegraphics[width=1.0\linewidth]{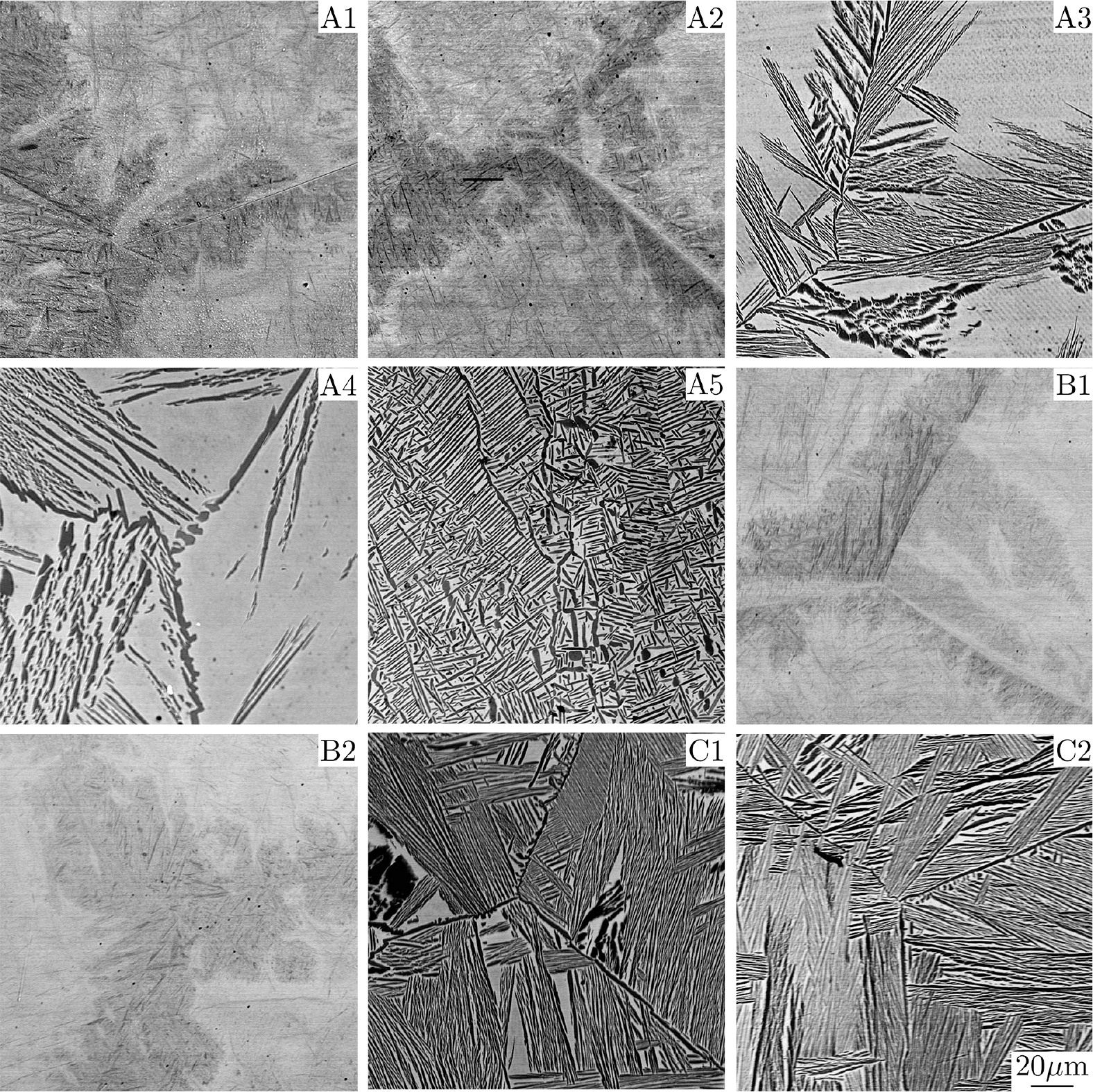}
\caption{Backscattered electron images of $\alpha_p$ plate growth during isothermal growth using an electro-thermal mechanical tester with resistance heating. A1-A5, growth at $850\celsius$ at times of 5, 10, 15, 20 and $40\usk\minute$ respectively. B1-2, growth at $900\celsius$ for 10 and $20\usk\minute$. C1-2, growth at $800\celsius$ for 20 and $40\usk\minute$. }\label{fig:BSEM}
\end{figure}

The experimental validation of the model can now be considered. The observed transus was $940\celsius$, compared to a prediction of $954\celsius$. \textcolor{black}{Careful transus measurements typically show an uncertainty of $\sim5\celsius$; the difference is probably attributable to commercial heat-to-heat variability.}  The observed microstructures are shown in Figure~\ref{fig:BSEM}, showing typical Widmanst\"{a}tten plate lengths grown from the prior $\beta$ grain boundaries after different isothermal ageing times.

The equilibrium volume fraction eventually obtained of primary $\alpha$ at $800\celsius$ is significantly greater than at $900\celsius$. In these images, the $\alpha$ phase appears dark due to the atomic number contrast obtained in backscatter electron imaging.  At short times at $850\celsius$ and $900\celsius$ (A1-2, B1-2), it is difficult to see much contrast between the phases. This is due to the sample not reaching chemical equilibrium during time at temperature, as the velocity of growth is slower. All images were taken at a beam voltage of 8kV with the smallest possible working distance, to maximise resolution. This was particularly useful for imaging smaller $\alpha$ grains, and ensured the best possible image resolution, reducing some of the uncertainty in the $\alpha$ length measurements. 

EBSD measurements were challenging, because of the thinness of the $\alpha$ plates and the residual strain from the transformation, which was not relaxed due to the quenching applied. This was rectified by etching the samples in Kroll's solution ($100\usk\milli\litre$ H$_2$O, 6$\usk\milli\litre$ HNO$_3$, 3$\usk\milli\litre$ HF) for 5 seconds, then polishing in colloidal silica suspension (OPS) for a further 5 minutes to remove the heavily damaged layer, and to prevent shadowing due to topographical relief when collecting the EBSD maps. In the samples where this step did not improve the resulting patterns, EBSD was completed at a beam voltage of 30$\usk\kilo\volt$ to increase signal from the $\alpha$ phase. Using these steps allowed the selection of appropriate grains for measurement.

\begin{figure}[t!]
\centering\includegraphics[width=1\linewidth]{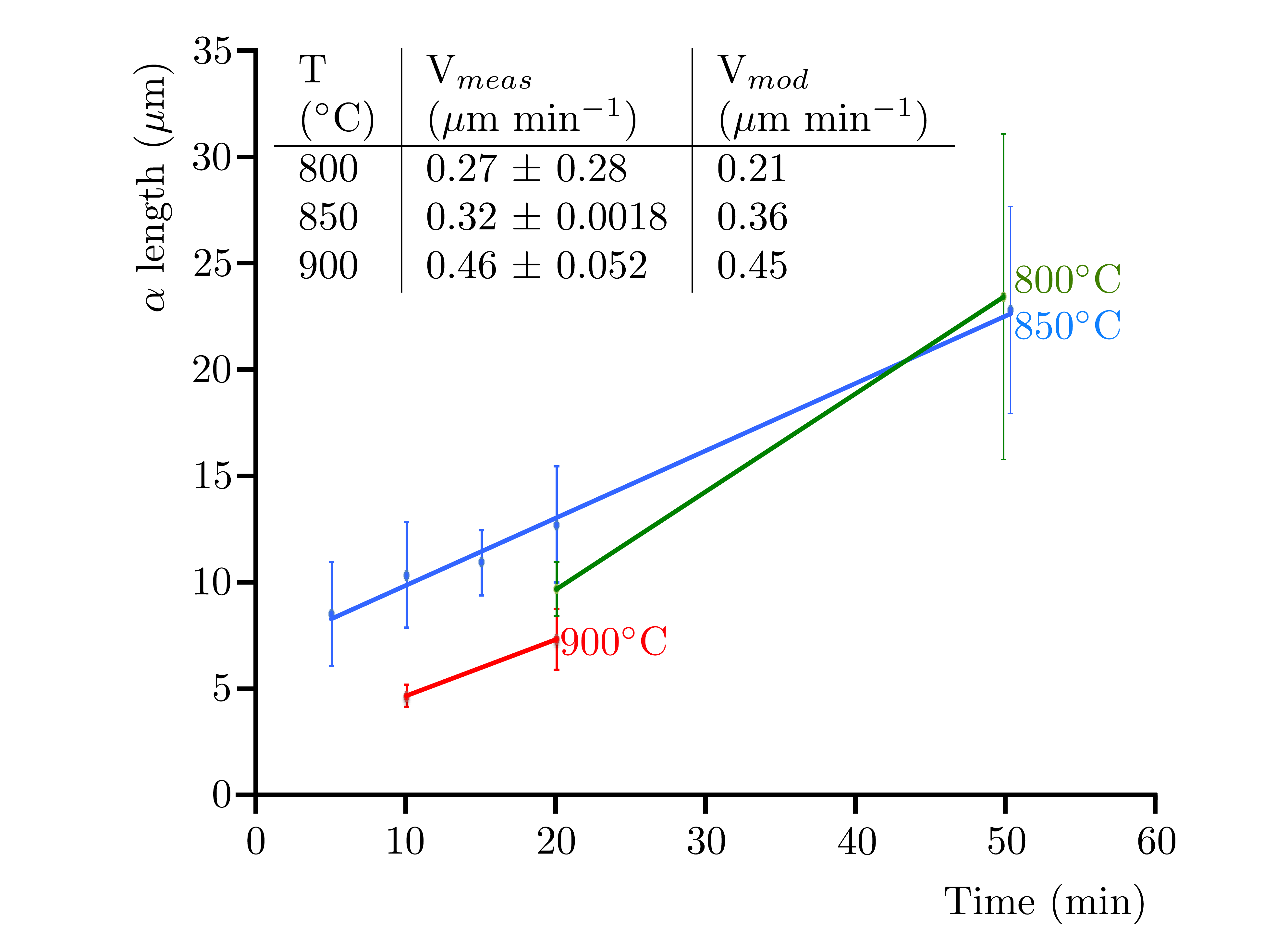}
\caption{Evolution of Widmanst\"{a}tten $\alpha$ plate length with temperature. Each measurement represents an average of a minimum of 20 plates, taken from at least two different prior $\beta$ grain boundaries in each sample, allowing uncertainties to be derived. The inferred straight line fits to the plate growth velocity are inset, along with the modelling results.}\label{fig:vels-meas}
\end{figure}

The evolution of plate length with temperature and the measured plate growth velocities are provided in Figure~\ref{fig:vels-meas}. It is found that the velocity increases with temperature, and that the growth rate is around $0.32${$\pm0.002$}$\usk\micro\meter\usk\minute^{-1}$ at $850\usk\celsius$. This agrees well with the model, which predicts $0.36\usk\micro\meter\usk\minute^{-1}$ at this temperature, which is very encouraging. All three model predictions are within 20\% of the measured velocities. It should be noted that experimental validation was not possible at the highest temperatures close to the transus, where the growth rate would be expected to drop. 

Interestingly, the data imply that there is an initial burst of plate growth as the diffusion field is established before steady-state is reached, and that this effect is reduced at higher temperatures. A similar effect is often observed in the analogous problem of dendrite growth during solidification using \emph{in situ} solidification radiography~\cite{Prasad2015,Xu2018}, and is a consequence of nucleation undercooling. It is also observed that a grain boundary $\alpha$ film is present in the samples cooled to $850\usk\celsius$, which is thinner in the $\alpha$ grown at higher temperatures. Therefore, it is implied that thicker grain boundary $\alpha$ films provide better nucleation sites for initial colony $\alpha$ growth.

The model results are strongly dependent on the diffusivity assumed. The diffusivity data available relate to electron probe microanalysis and metallographic measurements that date to 1963~\cite{Gibbs1963a,Elliot1962}, with $D_0=1.23\times10^{-8}\usk\meter^2\usk\second^{-1}$ and $Q/R=16,600\usk\kelvin$, and relate to the isolated $\beta$ phase at temperatures in excess of $1000\celsius$. Their extrapolation to low temperatures in the $\alpha+\beta$ phase field therefore results in significant differences between the measurements, over $3\times$, at $850\usk\celsius$. The interface solubility in the $\beta$ is probably reasonably well described by thermodynamic databases, as these are calibrated against experiment, whereas the plate widths $R$ observed can be quite variable. Therefore approximate agreement is probably the best that can reasonably be expected given the uncertainties in the experimental data; this argues strongly that future effort should be directed at improving the diffusivity data before more elaborate modelling approaches such as phase-field modelling could be applied to provide improved fidelity.

\subsection{Mo Diffusion around Growing Plates}
\begin{figure}[t!]
\centering\includegraphics[width=0.7\linewidth]{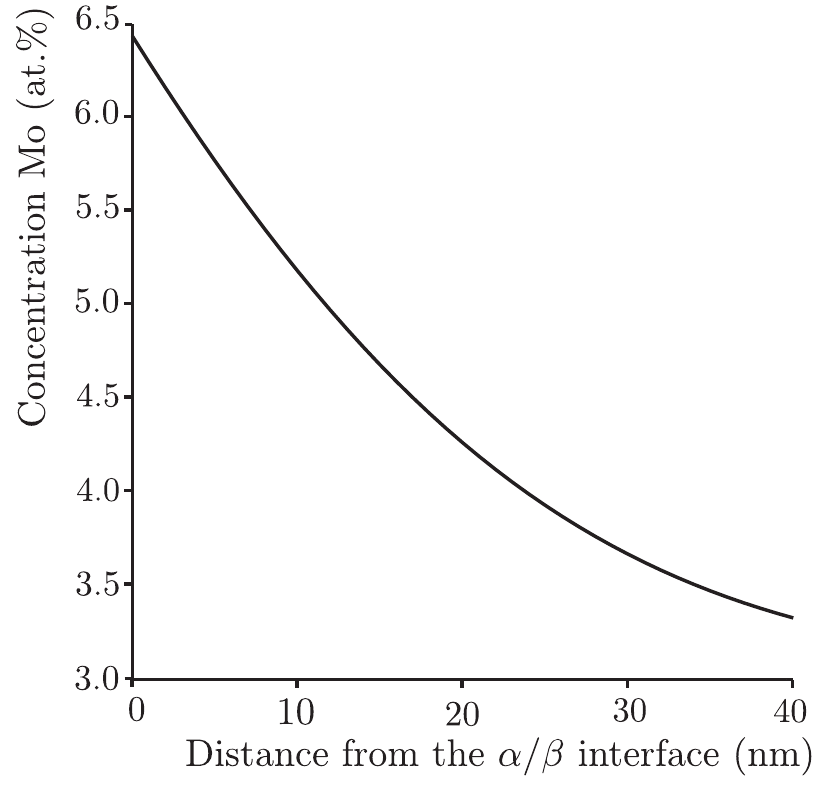}
\caption{Diffusion of molybdenum away from the $\alpha$/$\beta$ interface, for $\alpha$ plate growth at $800\celsius$ using the predicted phase solubilities from Pandat. }\label{fig:hill}
\end{figure}

A further source of validation of the model can be found from examination of the solute pile-up around the growing $\alpha$ plates. The predicted diffusion profile at $800\usk\celsius$ \textcolor{black}{after $30\usk$min} is shown in Figure~\ref{fig:hill}. As Mo is a $\beta$ stabiliser, it is rejected into the surrounding matrix as the $\alpha$ grows. Mo is the slowest diffusing element in the alloy system~\cite{Elliot1962,Semiatin1996}, and therefore is the rate-limiting species. At this temperature, $C_0$ (the interface solubility from Pandat) is 6.4 at\%, whilst the far-field composition $C_\beta$ is the alloy composition, 6 wt.\% or 3.0 at.\%.

\begin{figure}[t!]
\centering\includegraphics[width=1\linewidth]{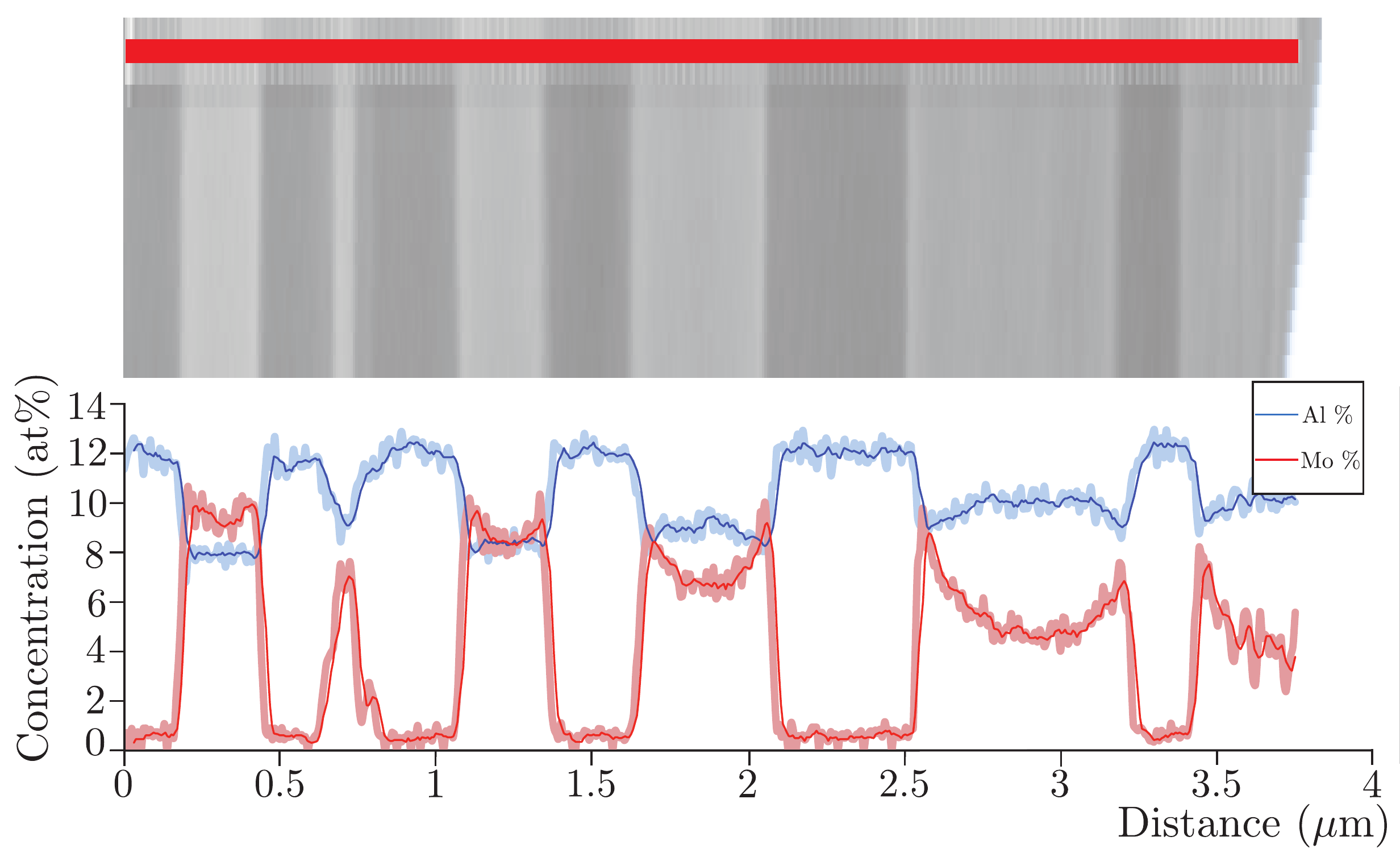}
\caption{Measurements taken using STEM-EDX on a specimen that has been heated to $960\celsius$, cooled at $7\celsius\usk\minute$$^{-1}$ to $800\celsius$ and water quenched manually. The solid red line on the HAADF-STEM image shows where the data was collected during the line scan, with the $\alpha$ phase appearing as 6 dark grey plates and the $\beta$ phase appearing as light grey. The TEM foil was prepared so the $\alpha$ is (0002) and the $\beta$ is (111). The blue data represents aluminium and the red data molybdenum. The lighter colour denotes the raw data, and the solid line shows a 5 point average.  The difference in the percentage of molybdenum can be seen between the two phases, with a higher intensity at the $\alpha$/$\beta$ interface.}\label{fig:STEM}
\end{figure}

A high angle annular dark field-STEM image (HAADF-STEM) of a series of plates is shown in Figure~\ref{fig:STEM}. It can be observed that the plate widths and spacings are quite heterogeneous, varying by more than a factor of three. \textcolor{black}{Due to the closeness of the primary $\alpha$ plates, there may be some interaction between the $\alpha$ plates.} The corresponding STEM-EDX linescan results are shown in Figure~\ref{fig:STEM}; as expected, Mo is rejected from the growing $\alpha$ plate and Al is absorbed. It can be observed that the interface concentration of Mo in the $\beta$ phase is quite variable, even for well separated plates. In addition, because the plate growth results in soft impingement of the diffusion fields, the far-field concentration does not always return to that of the alloy.

There are also fundamental measurement complications to consider in the quantitative as opposed to qualitative utilisation of STEM-EDX data. The measured average Al content is $9.4\usk$at.\%, whereas the alloy content is $10.8\usk$at.\%. Absolute composition measurement by STEM-EDX is complicated both by the specimen tilt towards the detector, specimen thickness and consequent absorbtion, and by the Cliff-Lorimer $k_{AB}$-factors used, which are species- and solution-dependent~\cite{Williams2009}. While these can be overcome through the use of appropriate standards, this is not commonly undertaken.  However, the data presented here allow a qualitative understanding of the spatial variation over $\micro\meter$ lengthscales to be developed. The concentrations in the growing $\alpha$ are, in comparison to the retained $\beta$, quite consistent between plates, as might be expected from thermodynamic considerations.

\begin{figure}[t!]
\centering\includegraphics[width=0.9\linewidth]{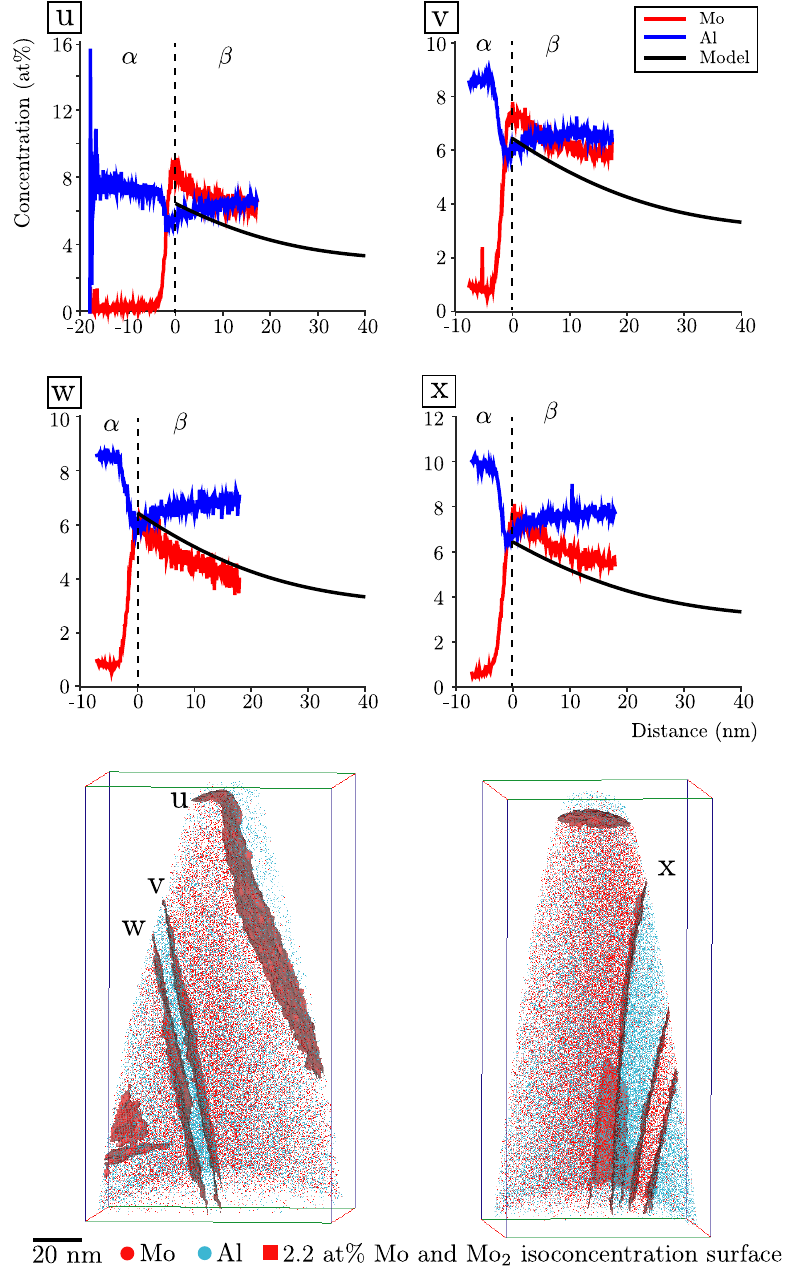}
\caption{\textcolor{black}{Atom probe tomography data from 2 different needles from a specimen that was heated to $960\celsius$, cooled at $7\celsius\usk\minute^{-1}$ to $800\celsius$ and water quenched. On the Al and Mo ion maps, isoconcentration surfaces at 2.2 at.\% Mo are shown to highlight the interfaces. Proxigrams were taken at interfaces x, u, w and v labelled on the ion maps. The Mo and Al concentration profiles obtained from interfaces u-W are shown and compared to the modelled Mo profiles.}}\label{fig:APprofiles}
\end{figure}

Therefore atom probe tomography was used to examine the interface composition profiles. Whilst there are evaporation and detection efficiency concerns in the atom probe, these are better controlled for quantitative comparison than the STEM-EDX data. The reconstruction of the \textcolor{black}{two needles} examined is shown in Figure~\ref{fig:APprofiles}. 


\textcolor{black}{Four interfaces (u-w) were found with sufficient diffusion distance of retained $\beta$ to give an undisturbed solute field; interfaces v and w from a thin plate completely contained in the APT needle, interface u from an $\alpha$ plate  at the top of the needle and interface $x$ from a plate with a neighbour on its other side. Proxigrams were taken at these four interfaces for comparison of the Mo solute field with the model. It can be seen that the interface concentration is found to vary, by up to $2\usk$at.\%, but the gradient of the solute field is reproduced. The interface concentration might be increased from that expected thermodynamically at $800\celsius$ if the plate initially began to form during cooling from above the transus, and might be decreased if plate growth ceased (e.g. due to soft impingement), allowing the solute peak to diffuse away.  It should also be noted that, despite the reasonable agreement obtained, the planar growth model used is not completely representative of the growth of a plate that also possesses a growing tip.}


\textcolor{black}{Close inspection of Figures 10-11 indicate that the primary $\alpha$ plate widths show substantial variability, with the 6 plates studied in Fig 10 being $<0.5\usk\micro\meter$ in width. The two completely enclosed in the two atom probe needles had to be chosen to be smaller in order to be contained in the needle. It should be noted that the plate radius $R=0.375\usk\micro\meter$ chosen in the modelling was determined based on the SEM imaging of a far larger number of plates; this difference should also be recalled when comparing the measured and modelled diffusion fields.}




\section{Discussion}
The model presented here can be applied to continuous cooling, as applied industrially to large closed-die forgings, through the use of Grong's isokinetic approach ~\cite{Grong2002,Myhr1991}. Here, individual intervals of growth at each temperature during cooling are numerically integrated, ignoring any second-order effects of the differences in the diffusion fields around the growing $\alpha$ plate. This approach is applicable to both dissolution and precipitation reactions. \textcolor{black}{The effect of the growing radius tip on the changing thickness of an $\alpha$ lamellae is not investigated here. There is a clear relationship between the two, as mathematically investigated by Hoyt~\cite{Hoyt2013}. For a steady state growth velocity, the relationship between the radius of curvature and the thickness of the growing lamellae must be upheld. There may also be and effect of additional $\alpha$ plates, causing impingement, which would additionally effect the growth kinetics.}

\begin{figure}[t]
\centering\includegraphics[width=90mm]{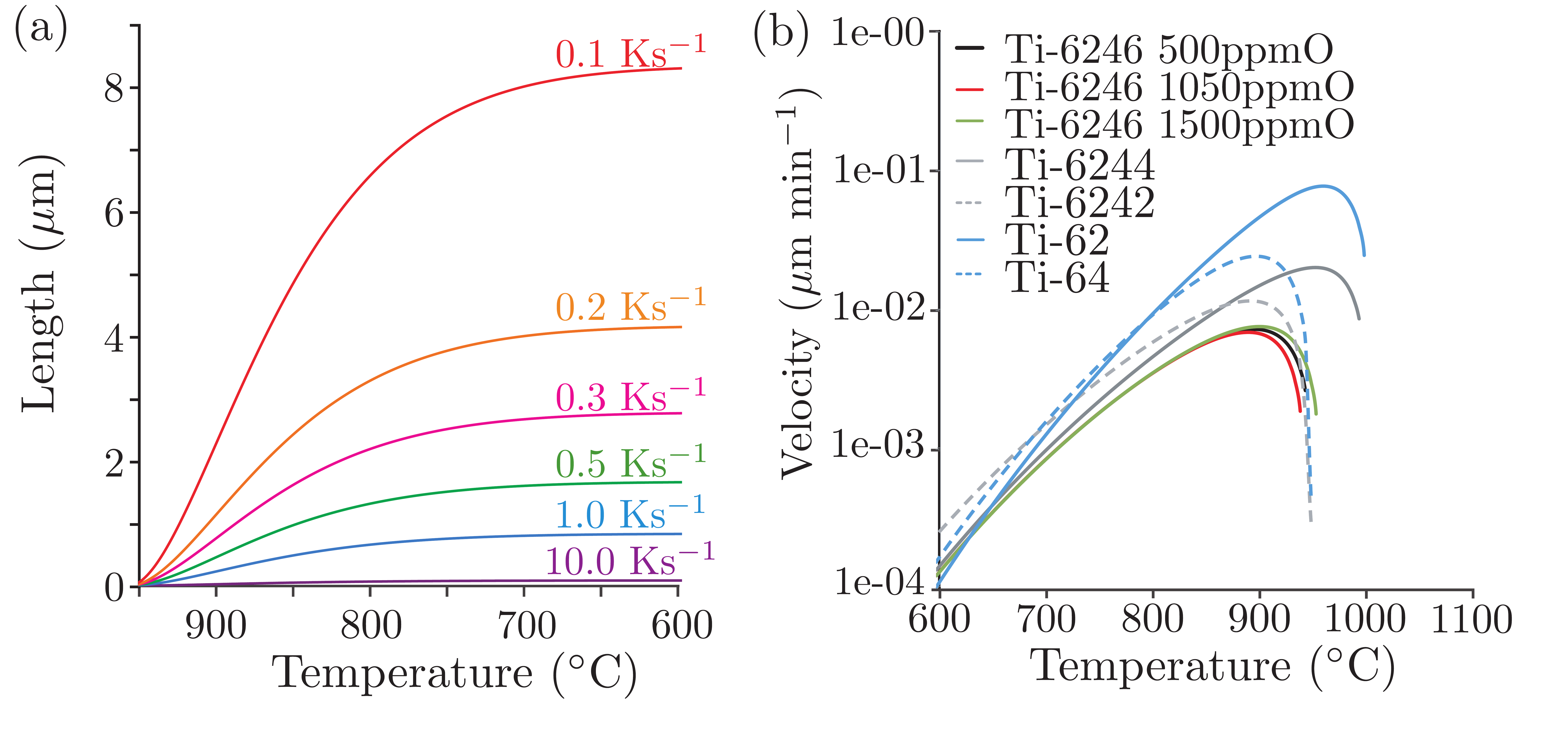}
\caption{(a) Effect of cooling rate on the plate length with temperature during cooling at constant cooling rates, in Ti-6246 with 1050 ppmwO. (b) plate growth velocity dependence on temperature for the alloys in Table~\ref{tab:alloy}.}\label{fig:ISKmodel}
\end{figure}

Figure~\ref{fig:ISKmodel}(a) and Table~\ref{tab:alloy} show the predicted plate length with temperature during cooling at different cooling rates. It is predicted that high cooling rates provide little opportunity for plate growth before diffusion becomes negligible and the phase transformation stops. 

\begin{table*}[t!]
\begin{center}\begin{small}
\begin{tabular}{l|ccccc|cccccc}
Alloy & $T_\beta$  & f $\alpha$ & $V$                            & $P(\Omega)$ & $D\usk\times10^{15}$ & \multicolumn{6}{c}{Final plate length}\\
wt.\% & $\celsius$ & mol.\%     & $\micro\meter\usk\minute^{-1}$ & (-)          & $\meter^{2}\usk\second^{-1}$ & \multicolumn{6}{c}{$\micro\meter$} \\\hline
Cooling rate (K$\second^{-1}$) &&&&&& 10 & 1 & 0.5 & 0.3 & 0.2 & 0.1\\\hline
Ti-6246-0.05O & 946 & 88 & 0.43 & 0.16 & 9.09 & 0.1	&1.0 &2.1&3.4&5.2&10.3  \\
Ti-6246-0.105O  & 954 & 88 & 0.52 & 0.15 & 9.09 & 0.1&1.1&2.2&3.7&5.5&11.1\\
Ti-6246-0.15O & 961 & 88 & 0.47 & 0.36& 9.09& 0.1&1.2&2.4&3.9&5.9&11.8\\
 Ti-624-2Mo  & 1004 & 92 & 0.92 & 0.31 & 9.09 & 0.3&3.0&6.0&9.9&14.9&29.8 \\
Ti-624-4Mo  & 978 & 95 & 0.67 & 0.23 & 9.09&0.2&1.8&3.5&5.9&8.8&17.6 \\
Ti-6Al-2V & 1006 & 95 & 2.82 & 0.36 & 24.2& 1&10&19&32&48&95 \\
Ti-6Al-4V & 980 & 87 & 1.32 & 0.23 & 24.2 & 0.3&3.2&6.4&10.7&16.0&32.0\\
\end{tabular}\end{small}\end{center}
\caption{Effect of alloying on the predicted transus temperature $T_\beta$, mol. fraction $\alpha$ at $600\usk\celsius$; and velocity $V$, $P(\Omega)$ and diffusivity $D$ at $900\usk\celsius$, for Mo or V. Unless stated, the alloys considered contained 1050 ppmw O (Ti-624$x$Mo) and 1500 ppmw O (Ti-6Al-$x$V). The final isokinetic model plate lengths are tabulated for the different cooling rates between 0.1 and $10\usk\kelvin\usk\second^{-1}$.}\label{tab:alloy}
\end{table*}

Figure~\ref{fig:ISKmodel}(b) shows the effect of O content, which alters the $\beta$ transus, Mo content, which changes both the transus and growth kinetics, and examines Ti-6Al-$x$V for comparison with the Ti-624-$x$Mo alloys. In all the alloys the velocity peak is around $40\usk\celsius$ below the transus. Increasing the O content, which will frequently be performed in order to provide solid solution strengthening, increases the $\beta$ transus by$\usk8\usk\kelvin$ per 500 ppmw O and therefore provides the opportunity for additional plate growth at high temperatures. However, this does not substantially affect the overall plate lengths achieved, because this is far from the maxima in growth rate due to the low supersaturation $\Omega$. 

In contrast, increasing the Mo content of Ti-624-$x$Mo alloys containing $1050\usk$ppmw O, from $x=2$ to 6, affects the transus by around $25\usk\kelvin$ per 2 wt.\% Mo; this will have a small effect on the final plate lengths obtained. However, the effect of Mo content on the supersaturations $\Omega$ obtained results in far larger changes in $V=2D(T)P(\Omega)/R$ and therefore in the final plate length, which is reduced by two-thirds for every 2 wt.\% Mo added. 

A third compositional effect that can be examined is to compare alloys where the $\beta$ is primarily stabilised by Mo as the slowest diffusing species to those where V is the slowest diffusing species, as in Ti-6Al-4V. It is observed that the growth rates in V-containing alloys are much faster than in Mo-stabilised alloys.  In terms of Mo-equivalence, which refers to the $\beta$ stabilising effect of different elements, V is two-thirds as effective as Mo (in wt.\%)~\cite{Titanium}, so 4V is equivalent to 2.7Mo. However, because of its higher diffusivity and the shape of the phase diagrams, a 4V containing alloy would possess faster plate growth kinetics. Of particular note is the very lightly $\beta$ stabilised alloy with just 2V, which would be expected to experience very fast plate growth.

These results are of utility to the alloy developer and heat treatment designer, where in general the goal would be to fill the prior-$\beta$ grain with colony $\alpha$, whilst leaving enough supersaturation of $\alpha$-stabilisers in the beta to allow for age-hardening through the use of secondary $\alpha$. This is of industrial relevance as it allows determination of the maximum cooling rate that can applied if the interior of the largest $\beta$ grain is to be safely reached by the Widmanst\"{a}tten $\alpha$ growing from the grain boundary.  Therefore the distances found in Figure~\ref{fig:ISKmodel} can be treated as the maximum $\beta$ grain size that can be tolerated in the forging for a given cooling rate. In addition, as larger ingot oxygen content will give rise to a higher transus temperature $T_\beta$, this means that elevated O contents will give rise to higher cooling rates that can be used, or conversely larger prior $\beta$ grain sizes that can be tolerated; these are also depicted in Figure~\ref{fig:ISKmodel}; however it should be cautioned that O effects are not large.

In the framework of Integrated Computational Materials Engineering (ICME), these results will be of utility to alloy designers and processing engineers interested, for example, in developing alloys where the prior $\beta$ grain can be filled with colony $\alpha$ whilst leaving a reservoir of supersaturated $\beta$ from which secondary $\alpha$ can be precipitated on ageing. 

\section{Conclusions}
A model for primary $\alpha$ plate growth in titanium alloys has been presented and validated experimentally compared to both observed plate growth rates  and the diffusion field observed by STEM-EDX and atom probe tomography. The following conclusions can be drawn.

 1. Plate growth velocity depends sensitively both on the diffusivity $D(T)$ and therefore on the rate-limiting species, and on the Mo content through its effect on the supersaturation $\Omega$, which controls the dimensionless velocity or Peclet number $P(\Omega)$. Therefore a maximum in plate growth velocities around $40\usk\kelvin$ below the transus is predicted.

 2. At $850\usk\celsius$ the plate growth velocity was found to be around $0.32\usk\micro\meter\usk\minute^{-1}$, which was in good agreement with the model prediction of $0.36\usk\micro\meter\usk\minute^{-1}$, although this is dependent on the diffusivity of the rate limiting species in the $\beta$ and on the plate width, neither of which are especially well known.

 3. The solute field around the growing plates, and the plate thickness, was found to be quite variable, as a result of the intergrowth of plates and soft impingement. Nevertheless, the solute field found around plates grown at $800\usk\celsius$ was found to extend to up to $\sim30\usk\nano\meter$, and the interface concentration in the $\beta$ was found to be around $6.4\usk$ at.\% Mo. This was in good agreement with the thermodynamic and kinetic modelling predictions.

 4. The sensitivity of the plate growth kinetics to alloy composition was examined. It was found that increasing O content had little effect the plate lengths expected during continuous cooling, through the effect on the solvus temperature. In contrast, reductions in Mo content in Ti624-$x$ alloys, and V content in Ti-6Al-$x$V alloys, resulted in substantial increases in plate growth kinetics, due to the effect on the supersaturations available. Alloys using V as the $\beta$ stabiliser instead of Mo are expected to have much faster plate growth kinetics.

In the framework of Integrated Computational Materials Engineering (ICME), these results provide a useful tool for the integrated design of alloys and process routes to achieve tailored microstructures to maximise, \emph{e.g.} high cycle fatigue strength.

\section*{Acknowledgements}
\noindent Funding from EPSRC (grants EP/K034332/1, EP/M506345/1 and EP/M022803/1) are acknowledged; the technical assistance of Ben Wood, T Ben Britton, Vivian Tong, Ecaterina Ware and Mahmoud Ardkani was gratefully received. Chris Gourlay also assisted with useful conversations on the analogy with dendrite growth during solidification.


\end{document}